\documentclass[aps, prd, twocolumn, superscriptaddress, nofootinbib]{revtex4}
\usepackage{graphicx}
\usepackage{dcolumn}
\usepackage{amssymb,amsmath,amsthm,mathrsfs}

\begin{document}

\newcommand{\Eq}[1]{Eq. \ref{eqn:#1}}
\newcommand{\Fig}[1]{Fig. \ref{fig:#1}}
\newcommand{\Sec}[1]{Sec. \ref{sec:#1}}

\newcommand{\PHI}{\phi}
\newcommand{\vect}[1]{\mathbf{#1}}
\newcommand{\Del}{\nabla}
\newcommand{\unit}[1]{\mathrm{#1}}
\newcommand{\x}{\vect{x}}
\newcommand{\ScS}{\scriptstyle}
\newcommand{\ScScS}{\scriptscriptstyle}
\newcommand{\xplus}[1]{\vect{x}\!\ScScS{+}\!\ScS\vect{#1}}
\newcommand{\xminus}[1]{\vect{x}\!\ScScS{-}\!\ScS\vect{#1}}
\newcommand{\diff}{\mathrm{d}}

\newcommand{\be}{\begin{equation}}
\newcommand{\ee}{\end{equation}}
\newcommand{\bea}{\begin{eqnarray}}
\newcommand{\eea}{\end{eqnarray}}
\newcommand{\vu}{{\mathbf u}}
\newcommand{\ve}{{\mathbf e}}
\newcommand{\vk}{{\mathbf k}}
\newcommand{\vx}{{\mathbf x}}
\newcommand{\vy}{{\mathbf y}}

\newcommand{\uden}{\underset{\widetilde{}}}
\newcommand{\den}{\overset{\widetilde{}}}
\newcommand{\denep}{\underset{\widetilde{}}{\epsilon}}

\newcommand{\nn}{\nonumber \\}
\newcommand{\dd}{\diff}
\newcommand{\fr}{\frac}
\newcommand{\del}{\partial}
\newcommand{\eps}{\epsilon}
\newcommand\CS{\mathcal{C}}

\def\be{\begin{equation}}
\def\ee{\end{equation}}
\def\ben{\begin{equation*}}
\def\een{\end{equation*}}
\def\bea{\begin{eqnarray}}
\def\eea{\end{eqnarray}}
\def\bal{\begin{align}}
\def\eal{\end{align}}


\title{Chiral vacuum fluctuations in quantum gravity}

\newcommand{\addressImperial}{Theoretical Physics, Blackett Laboratory, Imperial College, London, SW7 2BZ, United Kingdom}

\author{Jo\~{a}o Magueijo}
\affiliation{\addressImperial}
\author{Dionigi M. T. Benincasa}
\affiliation{\addressImperial}


\date{\today}

\begin{abstract}
We examine tensor perturbations around a deSitter background
within the framework of Ashtekar's variables and cousins
parameterized by the Immirzi parameter $\gamma$. At the classical
level we recover standard cosmological perturbation theory, with
illuminating insights. Quantization leads to real
novelties. In the low energy limit we find a second quantized theory of
gravitons which displays different vacuum fluctuations for right and
left gravitons. Nonetheless right and left gravitons have the same
(positive) energies, resolving a number of paradoxes
suggested in the literature. The right-left asymmetry of the
vacuum fluctuations depends on $\gamma$ and the ordering of the
Hamiltonian constraint, and it would leave a distinctive imprint in the
polarization of the cosmic microwave background, thus opening quantum
gravity to observational test.
\end{abstract}

 \pacs{To be done}

\maketitle


Loop quantum gravity is a promising scheme for quantizing the
gravitational field~\cite{ashbook,pulbook,rovbook,thbook}. At its
core lies the idea that the connection (or its holonomies), rather
than the metric, should be the central gravitational variable
driving quantization. This permits borrowing a number of
non-perturbative quantization techniques from non-abelian gauge
theories, notably the Wilson loop. The menace of
non-renormalizability can then be skirted, leading to a finite
theory. Regrettably the end product is far-removed from the real
world, with familiar concepts (such as smooth manifolds or
gravitons) finding no semi-classical niche in the theory, which
favors more abstract constructions like spin-networks. A
re-examination of the theory from the perturbative view-point is
in order, to establish whether it makes more pedestrian physical sense.

There have been a number of attempts to stave off the above criticism.
Loop quantum cosmology is a semi-classical scheme for deriving
effective Hamiltonians~\cite{lqc}; however its links with the
parent theory can be flimsy. Graviton states (and their loop
representations) were identified early on in loop quantum
gravity~\cite{gravitons}, but this work contained
a number of technical deficiencies (spelled out in this letter).
More recently, following on
from~\cite{gravitons}, Smolin proposed that fluctuations around
the Kodama state (a well-known exact solution to
the theory~\cite{kodama}) could provide well defined representations for
gravitons in a deSitter background~\cite{Leeclaim}. Witten claimed
that such gravitons would be pathological because one of the
helicities would have negative energy~\cite{wittenym}. This was
allegedly disproved in~\cite{leelaur}, but again in the shadow of
technical errors.

In this letter we re-examine the perturbative status of loop
quantum gravity following a simple guiding principle: we never
stray far from standard cosmological perturbation
theory~\cite{muk,lid}.
Clearly, well established {\it classical}
results in cosmology must have exactly equivalent descriptions
in Ashtekar's formalism; if they don't something has gone
wrong. Furthermore the loop quantization procedure should be mapped, in
some approximation, onto the usual inflationary calculation of
tensor vacuum quantum fluctuations. If differences arise one should 
understand their origin, and decide ``who's at fault''.

Crucial to this exercise are the reality conditions that
supplement Ashtekar's formalism. In order for the central concept
of duality to apply to a Lorentzian signature the geometry must
be complexified. Additional constraints then ensure that
``on-shell'' the geometry is real. This is implemented by the
inner product with which the Hilbert space is endowed, and the
implicit selection of physical (i.e. normalizable) states. In this
letter we show that physically sensible results
can only be obtained if we {\it include in all
expansions both positive and negative frequencies}. These should be 
associated with
graviton and anti-graviton states, to be identified only after reality
conditions are imposed.

Once this simple point is recognized a number of mysteries evaporate.
We reproduce Witten's negative energy gravitons~\cite{wittenym},
originally derived for Yang-Mills theories only. For example, for the
self-dual (SD) connection we find that
right-helicity (R) positive-frequency (+) and left-helicity (L)
negative-frequency ($-$) modes have positive energy, whereas
R$-$ and L+  modes have negative energy. However
we discover that the
pathological modes are not normalizable under the inner product
representing the reality conditions. Therefore they don't belong to
the physical Hilbert space, and indeed these modes don't exist
classically, i.e. by evaluating the SD connection using the
equations of motion.

The only physical modes are the usual particles: right and left
handed gravitons with a positive energy spectrum, albeit described
chirally (the right graviton appears in the positive frequency of
the SD connection, the left in its negative frequency). But a
dramatic novelty appears. For a standard ordering
of the Hamiltonian constraint only the negative frequency needs to
be normal ordered. Thus a significant difference appears in the
inflationary calculation for tensor vacuum fluctuations, using the
SD connection: a (scale-invariant) spectrum is produced, but only for
left gravitons. No right gravitons are produced.

Had we employed the anti-SD (ASD) connection, the
description would be reversed, leading to vacuum fluctuations
containing only right handed gravitons. More generally, Ashetkar's
SD and ASD connections belong to a class of connections parametrised
by the Immirzi parameter, $\gamma$. They hail from a canonical
transformation applied to General Relativity, resulting in
equivalent classical descriptions, {\it but inequivalent quantum
theories}. The main result in this letter is a reflection of this
fact at the perturbative level. We predict a $\gamma$ dependent 
chirality in the gravitational wave background. 
The effects on the polarization of the cosmic microwave background 
are unique~\cite{TBpapers}, opening up the doors to an
observational test of quantum gravity.

As our starting point we take metric: \be\label{pertmet}
ds^2=a^2[-d\eta^2 +(\delta_{ij}+h_{ij})dx^idx^j] \ee where
$h_{ij}$ is a TT tensor. For definiteness the
background is deSitter  (i.e. $ a=- 1/H\eta$, with
$H^2=\Lambda/3$ and $\eta<0$) but what follows can be repeated with
other backgrounds, and perturbations around Minkowski space-time
can be recovered by setting $H=0$. With a set of
conventions fully spelled out in~\cite{next} (and
following~\cite{thbook}) the connection is given by
$A^i=\Gamma^i+\gamma \Gamma^{0i}$, with
$\Gamma^i=-\frac{1}{2}\epsilon^{ijk}\Gamma^{jk}$. Here $\gamma$ is
the Immirzi parameter, and $\gamma=\pm i$ for the SD/ASD
connection. We then solve for the background using the
Einstein-Cartan equations and expand the canonical variables
as: \bea A^i_a&=&\gamma Ha \delta^i_a + \frac{a^i_a}{a}\\
E^a_i&=&a^2\delta^a_i - a\delta e^a_i \label{pertE}\; ,\eea where
$E^a_i$ is the densitized inverse triad, canonically conjugate to
$A^i_a$. Throughout this paper we'll adopt the following
convention: we define $\delta e^i_a$ via the triad 
$e^i_a=a\delta^i_a+\delta e^i_a$; we then raise and lower indices
in all tensors with the Kronecker-$\delta$, possibly mixing group
and spatial indices. This simplifies the notation
and is unambiguous if it's understood that $\delta e$ is
originally the perturbation in the triad.
It turns out that $\delta e_{ij}$ is then  proportional
to the ``$v$'' variable beloved by cosmologists~\cite{muk,lid}.

We now come to an important technical point. As
in the usual cosmological treatment we subject the perturbations
to Fourier and polarization expansions; however the
Ashtekar formalism presents us with some subtleties. If
reality conditions are yet to be enforced there must be graviton and
anti-graviton modes, so it's essential not to forget the negative
frequencies in all expansions, and ensure that they are
initially independent of the positive frequencies.
Furthermore, for a clearer physical picture, it's convenient to
use the quantum field theory convention stipulating that for
free modes the spatial vector $\vk$ points in the direction of
propagation {\it for both positive and negative frequencies}. This
is a simple point, but spurious couplings between $\vk$ and $-\vk$
modes otherwise come about, e.g. reality conditions constrain
gravitons moving in opposite directions, which is 
physically nonsensical.

Bearing this is mind we adopt expansions: \bea \delta
e_{ij}&=&\int \frac{d^3 k}{(2\pi)^{\frac{3}{2}}} \sum_{r}
\epsilon^r_{ij}({\mathbf k}) {\tilde\Psi}_e(\vk,\eta)e_{r+}(\vk)
\nonumber\\
&&
+\epsilon^{r\star}_{ij}({\mathbf k}) {\tilde\Psi}_e^\star (\vk,\eta)
e^{\dagger}_{r-}(\vk)\nonumber\\
a_{ij}&=& \int \frac{d^3 k}{(2\pi)^{\frac{3}{2}}} \sum_{r}
\epsilon^r_{ij}({\mathbf k}) {\tilde\Psi}_{a}^{r+}(\vk,\eta)a_{r+}(\vk)
\nonumber\\
&& +\epsilon^{r\star}_{ij}({\mathbf k}) {\tilde\Psi}_{a}^{r-
\star} (\vk,\eta)a^{\dagger}_{r-}(\vk) \label{fourrier}\eea where,
in contrast with previous literature (e.g.~\cite{gravitons,leelaur}),
$e_{rp}$ and $a_{rp}$  have two indices: 
$r=\pm 1$ for right and left helicities, and 
$p$ for graviton ($p=1 $) and anti-graviton ($p=-1$)
modes. In a frame with direction $i=1$ aligned with $\vk$ the
polarization tensors are~\cite{mtw}:
\begin{eqnarray}
\epsilon^{(r)}_{ij}&=&\frac{1}{\sqrt2} \left(\begin{array}{ccc}
0&0&0\\
0&1&\pm i\\
0&\pm i&-1\end{array}\right)\; .
\end{eqnarray}
The base functions have form ${\tilde
\Psi}(\vk,\eta)=\Psi(k,\eta) e^{i\vk\cdot \vx} $ and we
impose boundary conditions $\Psi(k,\eta)\sim{e^{-i k\eta}}$ when
$|k\eta|\gg 1$ for both $+\vk$ and $-\vk$ directions ($k=|\vk|>0$
throughout this letter). Only then does $\vk$ point in the direction
of propagation, as required. This convention has the essential
advantage of identifying the proper physical polarization
(until we know in which sense the mode is moving we cannot
assign to it a physical polarization). The functions $\Psi_e$
and $\Psi_a$ can in principle be anything, with the amplitudes
$e_{rp}$ and $a_{rp}$ carrying the necessary time dependence.
We may choose $\Psi$ so that they carry the full time dependence.
Hamilton's equations then merely confirm that the amplitudes
are constant,  but ${\tilde\Psi}_a^{rp}$ should have both $r$ and $p$
dependence. In these expansions we have already selected the
physical degrees of freedom (i.e. the Gauss and diffeomorphism
constraints have been implemented).

In order to canonically quantize the theory we need
its Hamiltonian formulation. We'll do this in detail
elsewhere~\cite{next} but stress that we can
read off the answer from cosmological perturbation
theory~\cite{muk,lid}. Functions $\Psi_e$ satisfy the same
equation as the variable ``$v$'' used by cosmologists.
Therefore, in a deSitter background: \be\label{eqnpsie}
\Psi_e''+{\left(k^2-\frac{2}{\eta^2}\right)}\Psi_e=0 ,\ee where
$'$ denotes derivative with respect to conformal time. This has
solution: \be\label{psie} \Psi_e=\frac{e^{-ik \eta}}{2 \sqrt  { k} }
{\left( 1-\frac{i}{k\eta} \right)}\; ,\ee where the normalization
ensures that the amplitudes $e_{rp}$  become
annihilation operators upon quantization. In addition, connection and metric are
related by Cartan's torsion-free condition $T=d e + \Gamma\wedge e=0$,
solved by \bea
\delta\Gamma^{0 i}&=& \frac {1}{a}{\delta e'_{ij}} \, dx^j \\
\delta \Gamma^{ij}&=&-\frac{2}{a}\partial_{[k}\delta e_{i ] j}\,
dx^j\label{gammaij}\; . \eea With the conventions given above
the second of these equations implies $\delta \Gamma^{i}=\frac{1}{a}\epsilon
^{ijk}\partial_{j}\delta e_{kl}\, dx^l$, so that
\be\label{arealsp} a_{ij}=\epsilon_{ikl}\partial_k\delta
e_{lj}+\gamma\delta{e}'_{ij} \; .\ee Inserting decomposition
(\ref{fourrier}) into this expression and using relation
$\epsilon_{nij}\epsilon^{r}_{il}k_j= i r k \epsilon ^{r}_{nl} $ we
get: \be\label{psia} \Psi_a^{rp}=\gamma p \Psi'_e + r
k\Psi_e\; , \ee 
(we have assumed $a_{rp}=e_{rp}$).
Inside the horizon ($|k\eta|\gg1$) this has the important
implication that
$\Psi_a^{rp}=(r-ip\gamma)k\Psi_e$ leading to the result that the
SD connection ($\gamma=i$) is made up of the right
handed positive frequency of the graviton and the left handed
negative frequency of the anti-graviton. The ASD connection
contains the other degrees of freedom (this result
was derived long ago~\cite{ash0} but seems to have been forgotten
in all subsequent work). For other values of $\gamma$ this is
shared differently, and as the modes leave the horizon
($|k\eta|\sim 1$) the classification breaks down.

The theory can now be quantized from Poisson brackets $
\{A^i_a(\vx),E^b_j(\vy)\}=\gamma
l_P^2\delta^b_a\delta^i_j\delta(\vx-\vy)$. They imply commutation
relations for the perturbative variables: \be\label{unfixedcrs1}
[a^i_a(\vx),\delta e^b_j(\vy)]= -i\gamma
l_P^2\delta^b_a\delta^i_j\delta(\vx-\vy)\; . \ee These are valid
before the Gauss and vector constraints are enforced and must be
replaced by a TT projected $\delta$-function upon gauge fixing.
Once this is done (details to be presented in~\cite{next}, but
see~\cite{weinberg}) we have: \be\label{fixedcrs} [{\tilde
a}_{rp}(\vk),{\tilde e}_{sq}^\dagger(\vk ')] =-i\gamma p
\frac{l_P^2}{2}\delta_{rs}\delta_{p{\bar q}} \delta(\vk-\vk ')\; ,
\ee where ${\bar q}=-q$ and ${\tilde a}_{rp}= a_{rp}\Psi_a^{rp}$
and ${\tilde e}_{rp}=e_{rp}\Psi_e$. In addition we must fix the
inner product of the Hilbert space to implement the reality
conditions. The reality of the metric ($\delta e_{ij}=\delta
e_{ij}^\star$) implies $ \label{realg}e_{r+}(\vk)=e_{r-}(\vk) $
i.e. the graviton and anti-graviton are identified, polarization
by polarization, mode $\vk$ by mode $\vk$. This is eminently
sensible. Reality conditions should never relate different
polarizations, or modes $\vk$ and $-\vk$, since gravity waves are
real (even if a complex notation is used~\cite{mtw}). The presence
of such spurious couplings in the
literature~\cite{gravitons,leelaur} merely signals that the
direction of motion for a given mode was not properly
identified, and in consequence the polarization incorrectly
assigned. This is avoided by using expansions (\ref{fourrier}).

For the connection, the reality and torsion-free conditions are
combined: $a_{ij}$ is allowed to be complex but only to the
extent that's consistent with the metric being real, given the
torsion-free condition. However in the Hamiltonian formalism we
only need to impose $\Re A^i=\Gamma^i(E)$, leaving it for the
dynamics to discover that $\Im A^i=|\gamma|\Gamma^{0i}$. Thus, $
a_{ij}+{\overline a}_{ij}=2a \delta \Gamma_{ij}=2 \epsilon
^{ink}\partial_n \delta e_{k j}$, which in terms of expansion
(\ref{fourrier}) becomes: \be\label{real} {\tilde a}_{r+} (\vk,
\eta)+ {\tilde a}_{r-}(\vk,\eta) = 2 r k {\tilde
e}_{r+}(\vk,\eta)\; .\ee We defer the
reality conditions' implementation via the inner product until after we have
the Hamiltonian.

It is straightforward to repeat what follows for a
general $\gamma$, but for
clarity we'll make our point with $\gamma=\pm i$, which turn 
out to be the extreme cases. Then, 
the Hamiltonian is: \be {\cal H}=\frac{1}{2l_P^2}\int
d^3x N E^a_i E^b_j \epsilon_{ijk}(F^k_{ab}+H^2 \epsilon _{abc}
E^c_k)\; .  \ee Expanding, and keeping only second order terms
quadratic in first order perturbations leads to: \bea\label{ham21}
{^{2}_1{\cal H} }&=&\frac{1}{2l_P^2}\int d^3x [ -a_{ij}a_{ij} +2
\epsilon_{ijk}
\delta e_{li}\partial_j a_{kl} \nonumber\\
&& - 2 \gamma Ha \delta e_{ij}a_{ij} -2H^2 a^2 \delta e_{ij} \delta
e_{ij} ]\;. \eea
To this one must add the
boundary term:
$
{\cal H}_{BT}=-\frac{1}{l_P^2}\int d\Sigma_a N\epsilon_{ijk}
E^a_i E^b_j A_{bk}
$
which perturbatively becomes: $ {^{2}_1{\cal H}
}_{BT}=\frac{1}{l_P^2}\int d\Sigma_i \epsilon_{ijk}\delta e_{lj}
a_{lk}$. Writing it as the volume integral of
a divergence and adding it to (\ref{ham21}) produces:
\bea\label{effectham} {\cal H} _{eff}&=&\frac{1}{2l_P^2}\int d^3x
[ -a_{ij}a_{ij} -2 \epsilon_{ijk}
(\partial_j\delta e_{li}) a_{kl} \nonumber\\
&& - 2 \gamma Ha \delta e_{ij}a_{ij} -2H^2 a^2 \delta e_{ij} \delta
e_{ij} ]\; , \eea to be identified with the Hamiltonian of the
effective quantum field theory representing the
theory perturbatively. It's easy to see that 
``on-shell'' (i.e. using (\ref{arealsp})) this is the
stress-energy tensor of gravitational waves,
with the usual kinetic and gradient terms.

We proceed to find the quantum Hamiltonian for $k|\eta|\gg 1$.
We assume an $EEF$ ordering but what follows can be adapted to
other orderings. Inserting expansions (\ref{fourrier}) into
(\ref{effectham}) we find: \bea\label{hameff} {\cal
H}_{eff}&=&\frac{1}{l_P^2}\int d^3k\sum_r
g_{r-}(\vk)g_{r+}(-\vk)+g_{r-}(\vk)g_{r-}^\dagger(\vk)\nonumber\\
&+&g_{r+}^\dagger(\vk)g_{r+}(\vk)+
g_{r+}^\dagger(\vk)g_{r-}^\dagger(-\vk)\; , \eea
with:
\bea
g_{r+}(\vk)&=&{\tilde a}_{r+}(\vk)\\
g^\dagger_{r+}(\vk)&=&-{\tilde a}^\dagger_{r-}(\vk) +2kr {\tilde e}_{r-}^\dagger(\vk)\\
g_{r-}(\vk)&=&-{\tilde a}_{r+}(\vk) +2kr {\tilde e}_{r+}(\vk)\\
g_{r-}^\dagger(\vk)&=&{\tilde a}^\dagger_{r-}(\vk) \eea where we used
$\epsilon^r_{ij}(\vk)\epsilon^{s\star}_{ij}(\vk)=2\delta^{rs}$ (note
that with our conventions
$\epsilon_{ij}^r(-\vk)=\epsilon^{r\star}_{ij} (\vk)$). We have
identified (anti)-graviton creation and annihilation operators,
$g_{rp}^\dagger$ and $g_{rp}$, as
in~\cite{gravitons}. From (\ref{fixedcrs}) they inherit 
algebra: \be\label{galgebra}
[g_{rp}(\vk),g^\dagger_{sq}(\vk')]=-i\gamma l_P^2 
(pr)k \delta_{rs}\delta_{pq} \delta(\vk-\vk')\; . \ee 
As in~\cite{wittenym}, half the particles are found to
have negative energy (those with $i\gamma= pr$). 
The Hamiltonian also contains pathological particle production
terms: the first and last of (\ref{hameff}). These
features are removed once the inner product is defined.

Notice first that the reality conditions amount to demanding that
$g_{rp}^\dagger$ are indeed the hermitian conjugates of $g_{rp}$.
This fully fixes the inner product~\cite{tate,pulbook,gravitons}.
We work in a holomorphic representation for wavefunctions $\Phi$
which diagonalizes $g^\dagger_{rp}$, i.e.: $g^\dagger_{rp}\Phi(z)=
z_{rp}\Phi(z)$ ($z$ represents collectively all the
$z_{rp}(\vk)$). Then, (\ref{galgebra}) implies: \be
\label{grop} g_{rp}\Phi_{rp}=-i\gamma l_P^2 (pr) \frac{\partial
\Phi}{\partial z_{rp}} \; .\ee 
With ansatz ${\langle \Phi_1 | \Phi_2\rangle}=\int d z d {\bar
{z}} e^{\mu(z,{\bar z})} {\bar \Phi_1}({\bar z}) \Phi_2 (z)$,
condition ${\langle \Phi_1 |g_{rp}^\dagger
|\Phi_2\rangle}={\overline {\langle \Phi_2 | g_{rp}
|\Psi_1\rangle}}$ therefore requires: \be \mu(z,{\bar z})=\int
d{\vk}\sum_{rp}\frac{pr}{i\gamma l_P^2}z_{rp}(\vk){\bar z}
_{rp}(\vk)\; , \ee 
fixing ${\langle \Phi_1 | \Phi_2\rangle}$.
Integrating $g_{rp}\Phi_0=0$ leads to
vacuum $\Phi_0={\langle z|0\rangle}=1$. Particle states
are monomials in the respective variables, $\Phi_n={\langle
z|n\rangle}\propto (g_{rp}^\dagger)^n \Psi_0= z_{rp}^n$.  With the inner 
product just derived these
aren't normalizable for $i\gamma= pr$. Therefore such modes
should be excluded from the physical Hilbert space, and this
removes all pathologies found in the Hamiltonian. We stress 
that the quantum modes we have disqualified 
don't exist classically (see discussion after
(\ref{psia})). For example for $\gamma=i$ the only physical modes
are $G_R=g_{R+}$ and $G_L=g_{L-}$.

We therefore regain the usual physical Hamiltonian but with one
major difference. For $\gamma=i$, for example, ${\cal
H}^{phy}_{eff}\approx \frac{1}{l_P^2}\int d\vk \,  k
({{G}_{L}{G}_{L}}^\dagger + {{G}_{R}}^\dagger{G}_{R}) $ and so only
the left handed graviton needs to be normal ordered. Following
the standard inflationary calculation (extrapolating the 
vacuum expectation value $V_r$ of a mode from
$|k\eta|\gg 1$ to $|k\eta|\ll 1$) we discover a scale invariant 
spectrum {\it with left gravitons only}. 
Repeating this calculation (\cite{next}) for general
$\gamma$ shows that: \be {\cal H}^{phy}_{eff}\approx
\frac{1}{2l_P^2}\int d\vk \sum_r  k [{G}_{r}{
G}^\dagger_{r} (1+ir\gamma)+ {G}_{r}^\dagger{G}_{r}
(1-ir\gamma)]\nonumber \ee 
so, after normal ordering, right and left particles are exactly symmetric, but
a chiral $V_r$ is found with:
\be\label{chiraleq}
\frac{V_R-V_L}{V_R+V_L}=i\gamma\; . \ee 
Strictly speaking this calculation only covers imaginary
$\gamma$ in the range $-i<\gamma <i$, but an extension for all
$\gamma$ (including real) will be presented elsewhere~\cite{next}.
For standard Palatini gravity $\gamma=0$ and no effect is predicted.

In a longer paper~\cite{next} we'll spell out the various steps of
this calculation and generalize its scope. The relation between the
ground state defined above, its inner product, and the
perturbed Kodama state~\cite{kodama,leelaur} will also be
examined. 
A generalized formula (\ref{chiraleq}), combining $\gamma$ with the ordering
of the Hamiltonian constraint, will be presented (note that 
$FEE$ ordering reverses the above argument; $EFE$ ordering produces no
chirality at all).  In the meantime we have
shown how a perturbative re-examination of quantum gravity 
can be fruitful. We hope to have cleared up a few
misconceptions and paradoxes. Above all, we
derived a striking prediction for the theory, which
could be tested in upcoming CMB polarization experiments. 
There are other mechanisms to generate gravitational chirality 
(e.g.~\cite{steph,gianl,merc}), but the one pointed out in this letter 
is by far the simplest. As explained in~\cite{TBpapers}, even
moderate chirality in the gravitational wave background would render its
detection easier. ``Killing two birds with one stone''
was the expression used in~\cite{TBpapers} to qualify the ensuing 
state of affairs.

{\bf Acknowledgements}
We thank M. Bojowald, G. Calcagni, J. Halliwell, C. Isham, H.
Nicolai, L. Smolin, J. Sonner and K. Stelle for advice.


\end{document}